\documentclass[a4paper,12pt]{article}
\usepackage[numbers]{natbib}
\usepackage{amsthm}
\usepackage{amssymb} 
\usepackage{amsmath}
\usepackage{amsfonts}
\usepackage{verbatim}
\usepackage{stmaryrd} 
\usepackage[matrix,arrow]{xy}

\newlength{\ancho}
\textwidth=\ancho
\newcommand{\rel}[1]{\mathcal{R}\left(#1\right)}
\newcommand{\sem}[1]{\llbracket #1 \rrbracket}
\newcommand{\modal}[1]{\langle #1 \rangle}

\newcommand{\bisim}{\sim}
\newcommand{\iso}{\cong}
\renewcommand{\emptyset}{\raisebox{-0.17ex}{\large$\varnothing$}}
\newcommand{\Baire}{\mathcal{N}}
\newcommand{\calS}{\mathcal{S}}
\newcommand{\calU}{\mathcal{U}}
\newcommand{\calB}{\mathcal{B}}
\newcommand{\calL}{\mathcal{L}}
\newcommand{\Pow}{\mathsf{Pow}}

\newcommand{\comp}{\mathsf{c}}
\newcommand{\func}{\rightarrow}
\newcommand{\umu}{\underline{\raisebox{0ex}[0ex][0.07ex]{\hspace{-0.15em}$\mu$}\hspace{-0.1em}}}
\newcommand{\I}{\mathbb{I}}
\newcommand{\R}{\mathbb{R}}
\newcommand{\Q}{\mathbb{Q}}
\newcommand{\N}{\mathbb{N}}
\newcommand{\B}{\mathbf{B}}
\newcommand{\lmp}[1]{\mathbf{#1}}
\newcommand{\<}{\langle}
\renewcommand{\>}{\rangle}
\newcommand{\keywords}[1]{{\renewcommand{\thefootnote}{\relax}\footnotetext{\emph{Keywords:}
    #1}}}

\newcommand{\sii}{\Leftrightarrow}
\renewcommand{\phi}{\varphi}

\newcommand{\sig}{\ensuremath{\sigma}}

\newcommand{\Sig}{\ensuremath{\boldsymbol\Sigma}}
\newcommand{\bPi}{\ensuremath{\boldsymbol\Pi}}
\newcommand{\Del}{\ensuremath{\boldsymbol\Delta}}

\newcommand{\pca}{\Sig_2^1}

\newcommand{\om}{\ensuremath{\omega}}

\newtheorem{theorem}{Theorem}
\newtheorem{lemma}[theorem]{Lemma}
\newtheorem{prop}[theorem]{Proposition}
\newtheorem{corollary}[theorem]{Corollary}

\theoremstyle{definition}
\newtheorem{definition}[theorem]{Definition}
\newtheorem{example}{Example}
\theoremstyle{remark}
\newtheorem*{ack}{Acknowlegdements}
\DeclareMathSymbol{l}{\mathalpha}{letters}{"60}

\begin{document}
\title{Unprovability of the Logical Characterization of Bisimulation}
\author{Pedro S\'{a}nchez Terraf\thanks{Supported by CONICET and
    SECYT-UNC.}}
\maketitle
\begin{abstract}
We quickly review \emph{labelled Markov processes
  (LMP)} and provide a counterexample showing that in general
measurable spaces, event
bisimilarity and state bisimilarity differ in LMP. This shows that 
the logic in Desharnais~\cite{Desharnais} does not 
characterize state bisimulation in non-analytic measurable
spaces. Furthermore we show that, under current foundations of
Mathematics, such logical characterization is unprovable for spaces
that are projections of a coanalytic set. Underlying this construction
there is a proof that stationary Markov processes over general
measurable spaces do not have semi-pullbacks.
\end{abstract}

\section{Introduction}
One of the more interesting facts about the state of the art on Markov decision
processes over a continuous state-space is that there exist a number
of competing notions of \emph{bisimulation}. The essential difference
with the discrete case 
is the appearance of nonmeasurable sets, which e.g., inhibit the possibility
of extending straightforwardly Larsen and Skou \cite{LarsenSkou} notion
of probabilistic bisimulation.

To work in a concrete setting, we will use the framework of Labelled
Markov Processes (LMP). LMP have a labelled set of \emph{actions}
that encode interaction with the environment; thus LMP are a
reactive model in which there are different
transition subprobabilities for each action. The thesis \cite{Desharnais}
contains a thorough study and introduction to LMP.

The categorical approach to bisimulation is already present in Joyal et
al.\ \cite{Joyal} and was studied for LMP in \cite{DEP}. There,
the notion of \emph{zig-zag} morphism was defined and the relation of
\emph{bisimilarity} was given by a span of zig-zags. Zig-zags are exactly the
coalgebra morphisms for Giry's functor $\Pi$ \cite{Giry}. The major
obstacle  for this definition of bisimulation was that transitivity of
bisimilarity was proved by using structure results only available when
the state-space is analytic. This was done in \cite{DEP} by
using a technical result by Edalat \cite{Edalat} that constructed a
span of zig-zags given a cospan. To achieve this goal, Edalat established
explicitly the existence of regular conditional probability for the
universal completion of a Polish space. An alternative point of view,
restricted to the category of Polish spaces, can be found in Doberkat \cite{semiDoberkat}.

A Hennessy-Milner logic was also developed and in \cite{DEP} it was
proved that the relation of bisimilarity was characterized by this logic in
the case of an analytic state-space. Clearly, a notion of logical
equivalence must be transitive, so the problem of transitivity is more
general than that of the logical characterization of bisimilarity.

It was realized that a new notion of bisimulation was needed, and in
\cite{coco} Danos, Desharnais, Laviolette, and Panangaden defined
\emph{event bisimulation} in terms of  the measurable structure of LMP.
They proved that logical equivalence and event bisimilarity coincide
and that both can be phrased as a cospan of zig-zags. These results are in a
way consequences of the fact that cospans are far more easy to work with in a
coalgebraic setting.

In this paper, we construct a counterexample showing that in general
measurable spaces, event
bisimilarity and state bisimilarity differ in LMP. This shows that the
Hennessy-Milner logic used in \cite{DEP,Desharnais,coco} does not
characterize bisimulation in non-analytic measurable
spaces. The construction includes also a counterexample to the
existence of \emph{semi-pullbacks} in the category of stationary
Markov processes over general measurable spaces. 

The construction of our counterexample needs the existence of a nonmeasurable set. It is
known that it is consistent with current foundations of
Mathematics that there exists a nonmeasurable subset of the Euclidean
plane which is  the continuous image of the complement of an
analytic set. Hence, there are spaces in level 2 of the
\emph{projective hierarchy} of sets (level 0
occupied by Borel sets, level 1 by analytic sets and their
complements) for which the logical characterization bisimulation is
unprovable.

The paper is organized as follows. In Section~\ref{sec:background} we
review some background to our study, including some concepts related
to measurable spaces and prior results on
labelled Markov processes. Section~\ref{sec:extensions-measures}
develops the consequences of \L{}o\'s
and Marczewski's theorem on extension of measure, in
particular the non-existence of semi-pullbacks in the category of
stationary Markov processes and zig-zag morphisms. Our main
counterexample, a LMP for which event and state bisimilarity differ
from each other, is constructed in
Section~\ref{sec:counterexample}. A careful analysis of the set
theoretical requirements of this construction is pursued in 
Section~\ref{sec:furth-analys-constr}, where we show our unprovability
result.

\section{Background}\label{sec:background}
\subsection{Measurable spaces}
A \sig-algebra over a set $S$ is a family of subsets of $S$ closed
under countable union and complementation. Given an arbitrary family
$\calU$ of subsets of $S$, we use $\sig(\calU)$ to
denote the least \sig-algebra over $S$ containing $\calU$.

Let  $\<S,\calS\>$ a measurable
space, i.e., a set $S$ with a \sig-algebra $\calS$ over $S$. We say that  $\<S,\calS\>$ (or 
$\calS$) is \emph{countably generated} if there is some countable family
$\calU\subseteq\calS$ such that $\calS=\sig(\calU)$. Assume now that
$V\subset S$. We will use 
$\calS_V$ to denote  $\sig(\{V\}\cup\calS)$, \emph{the extension of $\calS$
by the set $V$}. It is immediate that
\[\calS_V = \{(B_1\cap V)\cup (B_2\cap V^\comp) : B_1,B_2\in\calS\}.\]
The \emph{sum} of two
  measurable spaces $\<S_1,\calS_1\>$  and $\<S_2,\calS_2\>$ is
  $\<S_1\oplus S_2,\calS_1\oplus\calS_2\>$, with the following  abuse
  of notation: $S_1\oplus S_2$ is the disjoint union (direct sum
  \emph{qua} sets) and $\calS_1\oplus\calS_2 = \{Q_1\oplus Q_2 :
  Q_i\in\calS_i\}$. We obtain:
\begin{equation}\label{eq:sum-spaces}
\<S,\calS_V\> \iso \<V,\calS|V\>\oplus  \<V^\comp,\calS|V^\comp\>.
\end{equation}
It is obvious that if $\calS$ is countably generated so is $\calS_V$. 

If $Y$ is a topological space, $\B(Y)$ will denote the \sig-algebra
generated by open sets in $Y$, hence $\<Y,\B(Y)\>$ is a measurable
space, the \emph{Borel space of Y}. 

The central example (see Theorem~\ref{th:iso-borel} below) is the
Borel space of the open unit interval $\I:=(0,1)$. The
\sig-algebra $\B(\I)$ is countably generated: it is generated by the
family  $\calB := \{B_a:a\in \om\}$ of all open subintervals of
$\I$ with rational endpoints. This 
family has a property which is inherited by the whole \sig-algebra: we
say that a family of sets $\calS\subseteq\Pow(S)$  \emph{separates
  points} if $x,y$ are distinct points in $S$, then there is
  $A\in\calS$ with $x\in A$ and  $y\notin A$. Hence $\B(\I)$ separates
points. We have the following propositions; the first of them is
immediate and we will use it without reference.
\begin{prop}
For $\calU\subseteq\Pow(S)$, $\calU$ separates points if and only if
$\sig(\calU)$ does.
\end{prop}
\begin{prop}[12.1 from~\cite{Kechris}]\label{p:sep-metriz}
The following are equivalent:
\begin{enumerate}
\item  $\<S,\calS\>$ is isomorphic to some $\<Y,\B(Y)\>$, where $Y$ is
  separable metrizable.
\item  $\<S,\calS\>$ is countably generated and separates
  points.
\end{enumerate}
\end{prop}
A topological space is \emph{Polish} if it is separable and completely
metrizable.  Examples of Polish spaces are the Euclidean spaces $\R^n$
and all countable discrete spaces. Polish spaces are closed under
countable product, and hence the \emph{Baire space} $\Baire := \N^\N$
is Polish, assuming $\N$ discrete. We have the following fundamental result (see \cite[15.6]{Kechris}):
\begin{theorem}[The Isomorphism Theorem]\label{th:iso-borel}
Let $Y$ be an uncountable Polish space. Then the Borel space
of $Y$  is isomorphic to $\<\I,\B(\I)\>$.
\end{theorem}
Finally, an \emph{analytic} (or $\Sig_1^1$) space is the continuous image of a Polish space.

%
\subsection{Labelled Markov Processes}
The following definitions are extracted from Danos et al.~\cite{coco}.

Let $\<S,\calS\>$ be a measurable space. Recall that a Markov kernel on a measurable space $\<S,\calS\>$ is a
function $\tau:S\times \calS \func [0,1]$ such 
that for each fixed $s \in S$, the set function $\tau(s, \cdot)$ is a
\mbox{(sub-)} probability measure, and for each fixed $X \in\calS$, $\tau(\cdot, X)$ is a $(\calS,\B([0,1]))$-measurable function.

Now let $L$ be any set.  
\begin{definition}\label{def:LMP}%
  A \emph{labelled Markov process (LMP)} is
  a structure $\lmp S = \<S,\calS, \{\tau_a  :  a\in L \}\>$ where $\<S,\calS\>$ is a
  measurable space  and for $a\in L$, $\tau_a : S \times \calS
  \rightarrow [0,1]$ is a Markov kernel. We will call $L$ the set of
  \emph{labels} and 
     $\<S,\calS\>$ the \emph{base space} of  $\lmp S$.
\end{definition}
Labelled Markov processes form a category whose arrows are given by  \emph{zig-zag
  morphisms}.
\begin{definition}
Let   $\lmp S = \<S,\calS, \{\tau_a : a\in L \}\>$ and  $\lmp S' =
\<S',\calS', \{\tau_a' : a\in L \}\>$. be LMP. A \emph{zig-zag
  morphism} $f:\lmp{S}\func\lmp{S'}$ is a surjective measurable map $f:
\<S,\calS\>\func\<S',\calS'\>$ such that for all $a\in L$ we have:
\[\forall s\in S \;\forall Q\in\calS' : \tau_a(s,f^{-1}(Q)) =
\tau_a'(f(s),Q).\]
\end{definition}
The reader may find variants of this definition along the
development of the theory. In \cite{Desharnais} LMP are augmented
with an initial state and zig-zags are not required to be surjective
but to preserve initial states. Later in \cite{DEP} the authors adopt
the present definition. However, these are minor differences. More
fundamentally, both \cite{Desharnais,DEP} require the state
space to be analytic.
We refer the reader to Desharnais \cite{Desharnais} for motivation and
for the fundamental results in the theory of LMP. 

Some notation concerning
binary relations will be needed to state the formal definitions. Let
$R$ a binary relation over $S$. 
A set $Q$ is \emph{$R$-closed} if $Q\ni x\mathrel{R} y$ implies $y\in
Q$. $\calS(R)$ is the \sig-algebra of $R$-closed sets in $\calS$. Lastly, 
let $\calU$ be a subset of $\Pow(S)$. The relation $\rel{\calU}$
is given by:
\[(s,t)\in\rel{\calU} \quad \iff \quad \forall Q\in \calU: s\in Q \sii
t\in Q.\]

Fix a LMP $\lmp S =
\<S,\calS, \{\tau_a : a\in L \}\>$. 
\begin{definition}
\begin{enumerate}
\item\label{def:lmp_sb}%
  A relation $R\subseteq S\times S$ is a \emph{state bisimulation}
  on $\lmp S$
  if it is symmetric and for all $a\in L$, $s\mathrel{R}t$ implies
  $\forall Q\in\calS(R) :
  \tau_a(s,Q) =  \tau_a(t,Q)$.
\item\label{def:lmp_eb}%
  An \emph{event bisimulation} on $\lmp S$  is a sub-$\sigma$-algebra
  $\calU$ of $\calS$ such that $\<S,\calU, \{\tau_a : a\in L \}\>$ is
  a LMP (i.e., $\tau_a$ is
  $\calU$-measurable for each $a\in L$).  We also say that a
relation $R$ is an event bisimulation if there is an event
bisimulation $\calU$ such that $R=\mathcal{R}(\calU)$.
\end{enumerate}
If there is a state (event) bisimulation $R$ such
$s\mathrel{R}t$, we will say that \emph{$s$ is state- (event-) bisimilar to $t$}. 
\end{definition}
It is proved \cite[Proposition 3.5.3]{Desharnais} that whenever there
exists a zig-zag morphism $f$ between two LMP $\lmp S$ and $\lmp T$,
the equivalence relation generated by the pairs $(s,f(s))$ with $s\in
S$ is a state bisimulation on the sum $\lmp S\oplus\lmp T$.%
\footnote{The base space of  the sum $\lmp S\oplus\lmp T$ is
  $\<S,\calS\>\oplus \<T,\mathcal{T}\>$ and the transition function
  $\tau_a^{\lmp S\oplus\lmp T} (r,A)$ equals $\tau_a^{\lmp S} (r,A\cap S)$
  if $r\in S$, and $\tau_a^{\lmp T} (r,A\cap T)$ if $r\in T$.}
       On the other hand, for every state bisimulation $R$ on a LMP
$\lmp S$, the identity map $Id : \<S,\calS, \{\tau_a : a\in L \}\> \to
\<S,\calS(R), \{\tau_a : a\in L \}\>$ is a zig-zag (see \cite[Lemma
  4.2]{coco}).

A generalization of the notion of event bisimulation will be needed in
the sequel:
\begin{definition}\label{def:stable+criterion}
A subfamily $\calU\subseteq\calS$ is \emph{stable} with respect to
$\lmp S$ if for all $A\in\calU$, $r\in [0,1]$ and $a\in L$, $\{s\in S
: \tau_a(s,A) >r \}\in \calU$.
\end{definition}
Since a function $f:S\func [0,1]$ is measurable if and only if
$f^{-1}((r,1])$ is a measurable set for every $r\in [0,1]$, an event
bisimulation on $\lmp S$ is the same thing as a stable
sub-\sig-algebra of $\calS$. This notion of stability was further
generalized by Doberkat \cite{Doberkat2007638} to the concept of
\emph{congruence} for stochastic  systems. 

It is shown that there exists a greatest state
bisimulation $\bisim$ (namely, the relation of \emph{state bisimilarity}), and in \cite{coco} it is proved that event
bisimulation is characterized by 
the  logic $\calL$ given by the following productions:
\begin{eqnarray*}
  \phi & \ \equiv \ &\textstyle
    \top \ \mid \ \phi_1\land\phi_2 \ \mid \ \modal{a}_q\psi
\end{eqnarray*}
where $a\in L$ and $q\in\Q\cap{[0,1]}$. 
Formulas in $\calL$ are interpreted as sets of states in which they
become true as follows:
\[ \sem{\top} := S \qquad  \sem{\phi_1\land\phi_2} :=
\sem{\phi_1}\cap\sem{\phi_2}  \qquad \sem{\modal{a}_q\psi} :=
\{s\in S : \tau_a(s,\sem{\psi})\geq q\}\]
Let $\sem{\calL} := \{\sem{\phi} : \phi \in \calL\}$. 
Two
states $s,t\in S$ are \emph{logically equivalent} if
$s\mathrel{\rel{\sem{\calL}}}t$, i.e., if they satisfy exactly the same
formulas. Given a class $\mathcal{M}$ of LMP, the problem of the
\emph{logical characterization of  bisimulation} for $\mathcal{M}$ is to prove the
following statement:  
\begin{quote}
  For all $\lmp S\in \mathcal{M}$  and all $s,t\in S$, 
  $s \mathrel{\rel{\sem{\calL}}} t$ if and only if 
  there exists a bisimulation $R$ such that $s \mathrel{R} t$. (We say that
  \emph{$\calL$ completely characterizes} bisimulation).%
\footnote{Perhaps a better phrasing would be ``characterization of
  \emph{bisimilarity}'', but we keep this one in accordance with
  previous works.}
\end{quote}
This depends on how do we qualify the word ``bisimulation''. In the
case of event bisimulation, we have the
following results:
\begin{theorem}[{\cite[Proposition 5.5]{coco}}]
 $\sig(\sem{\calL})$ is the smallest stable \sig-algebra.
\end{theorem}
\begin{theorem}[{\cite[Corollary 5.6]{coco}}]\label{thm:charaterization_event_bisimulation}%
  $\sig(\sem{\calL})$ is the least event bisimulation, and hence the
  logic $\calL$ completely characterizes event bisimulation. 
\end{theorem}
In view of this result we conclude that the problem of the logical
characterization of state bisimulation  is equivalent to  decide if
 event and state bisimilarity coincide. This  was also proved
 in \cite{coco}  for the class of LMP having an analytic base space.
 To obtain this result, one needs  the logic $\calL$ to be
countable, hence also limiting the set of labels $L$ to be at most
countable. It is noteworthy that the counterexample of
Section~\ref{sec:counterexample} conforms this restriction.
\section{Extensions of measures}\label{sec:extensions-measures}
The reader can consult Royden~\cite{Royden} and
Rudin \cite{Rudin} as general references for Measure Theory.

The key idea in the construction of our counterexample is the
possibility of extending the domain of definition of a (probability)
measure in a very flexible way. We will use a  result
due to \L{}o\'s
and Marczewski~\cite{Los} concerning \emph{canonical extensions} of
measures.%
\footnote{\L{}o\'s
and Marczewski use the term ``measure'' to mean a 
\emph{finitely} additive set function while reserving ``\sig-measure'' for a
standard (\sig-additive) measure. In any case, they prove the result
for both finitely and countably additive set functions.}%
 If $\calS\subseteq\calU$ and $\mu$,
$\nu$ are measures defined on  $\<S,\calS\>$,  $\<S,\calU\>$
(respectively), we say that $\nu$ \emph{extends $\mu$ to
  $\<S,\calU\>$} when $\nu | \calS = \mu$. We recall that the
 \emph{inner} and \emph{outer} measures defined from $\mu$, denoted
 $\mu_i$ and $\mu_e$ respectively, are the countably subadditive
 functions given by
\[\mu_i(A) := \sup \{\mu(M) : M\subseteq A, M\in\calS\} \quad \mu_e(A)
:= \inf \{\mu(M) : M\supseteq A, M\in\calS\},\]
for every $A\subseteq S$.

It is well known that the domain of definition of a measure $\mu$ can be
enlarged to include all subsets $A$ for which $\mu_i(A) =\mu_e(A)$;
such sets are called \emph{$\mu$-measurable} and they form a
\sig-algebra. In the case of Lebesgue measure, we will use the name
``Lebesgue measurable sets''.  By using the Axiom of Choice it can be
proved the existence of sets in Euclidean space that are not Lebesgue
measurable. For such  sets the following results are most significant.
\begin{theorem}
Let $\mu$ be a finite measure defined in $\<S,\calS\>$, and  let $V\subseteq
S$. Then $\umu$ and $\bar{\mu}$ defined as:
\begin{align*}
 \umu(E) &= \mu_i(E\cap V) + \mu_e(E\cap V^\comp)  \\
 \bar{\mu}(E) &= \mu_e(E\cap V) + \mu_i(E\cap V^\comp)  
\end{align*}
for every $E\in \calS_V$ are measures that extend $\mu$ to
$\<S,\calS_V\>$ and satisfy:
\[\umu(V) = \mu_i(V), \qquad  \bar{\mu}(V) = \mu_e(V).\]
\end{theorem}
\begin{proof}
By Theorems 4, 2, and 1 in \cite{Los}. The proof follows elementarily
from these facts:
\begin{enumerate}
\item For all $A,B$ such that $A\cap B=\emptyset$ and $A\cup B\in
  \calS$ we have $\mu_i(A)+\mu_e(B) =\mu(A+B)$ (see, for instance,
  \cite[14.H]{Halmos}).
\item For $E_j\in \calS_V$ ($j\in\om$) pairwise disjoint there are
  $M_j, N_j\in \calS$ such that $E_j = (M_j\cap V )\cup (N_j\cap
  V^\comp)$ and $M_j$ ($N_j$) pairwise disjoint.
\item If $A_j\subseteq M_j$ with $M_j\in\calS$ pairwise disjoint, then
  $\mu_i(\bigcup_j A_j) = \sum_j \mu_i(A_j)$ and   $\mu_e(\bigcup_j
  A_j) = \sum_j \mu_e(A_j)$.
\end{enumerate}
\end{proof}
\begin{corollary}\label{c:extension-by-nonmeas}
Let $\mu$ be a finite measure defined in $\<S,\calS\>$ and  let $V\subseteq
S$ be non $\mu$-measurable. Then there are extensions $\mu_1$ and $\mu_2$
to $\calS_V$ of $\mu$ such that $\mu_1(V)\neq\mu_2(V)$.
\end{corollary}
\begin{proof}
Immediate by definition of (non) $\mu$-measurable set.
\end{proof}
At this point it is possible to give a hint for the failure of the
logical characterization of bisimulation. The logic can be seen as an
encoding for the family $\sem{\calL}$ of measurable sets, which can be enlarged to the
\sig-algebra $\sig(\sem{\calL})$. This \sig-algebra cannot ``weigh''  a
set $V$ which is not measurable ``respect to $\sig(\sem{\calL})$'';
more precisely, one can have two measures that 
are equal on $\sig(\sem{\calL})$ (``logically equal'') but they differ on
$\sig(\sem{\calL})_V$.

With this tool at hand we are now ready to witness a failure for the existence
of semi-pullbacks  \cite{Edalat}
in the category of (labelled) Markov processes over general measurable spaces and
zig-zag morphisms. A category \emph{has semi-pullbacks} if for every
diagram consisting of 
objects $\lmp S_1$, $\lmp S_2$ and 
$\lmp T$ and arrows $f_i:\lmp S_i\func\lmp T$ ($i=1,2$; a \emph{cospan}) there exists
an object $\lmp S$ and arrows 
$\pi_i:\lmp S\func\lmp S_i$ (a \emph{span}) such that
$f_1\circ\pi_1=f_2\circ\pi_2$. Recall also that a \emph{stationary
  Markov process} \cite{Edalat} is a LMP with a single Markov kernel (i.e.,
the label set is a singleton). We have:
\begin{theorem}\label{t:no-semi-pullbacks}
  The category of stationary Markov processes and zig-zag morphisms does
  not have semi-pullbacks.
\end{theorem}
\begin{proof}
Let $m$ be Lebesgue measure on the closed interval $\bar\I := [0,1]$, let $\calS:=\B(\bar\I)$,
and let $V$ be a subset of 
$\bar\I$ that is not Lebesgue measurable. Take two extensions $m_0$ and
$m_1$ of $m$ to $\calS_V$ such that $m_0(V)\neq m_1(V)$ as in
Corollary~\ref{c:extension-by-nonmeas}. Let $\chi_M$ be the indicator
function of the set $M\subset\bar\I$. This function is
$\calS_V$-measurable if and only if $M\in\calS_V$.  Now define
\begin{align*}
\zeta(r,A) &:= \chi_{(0,1]}(r)\cdot  \delta_0(A) +\chi_{\{0\}}(r)\cdot m_0(A) \\
\theta(r,A) &:= \chi_{(0,1]}(r)\cdot  \delta_0(A) +\chi_{\{0\}}(r)\cdot m_1(A) \\
\tau(r,B) &:= \chi_{(0,1]}(r)\cdot  \delta_0(B) +\chi_{\{0\}}(r)\cdot m(B), 
\end{align*}
for every $0\leq r \leq 1$, $A\in \calS_V$ and $B\in \calS$. 
We will  prove that $\zeta$ is a Markov kernel over $\<\bar\I,
\calS_V\>$; the proofs for $\theta$ 
and that  $\tau$ is a Markov kernel over  $\<\bar\I, \calS\>$
are exactly analogous. To accomplish this, we have to check that $\zeta(r,\cdot)$ is a
(sub)probability measure for each $r\in \bar\I$, and $\zeta(\cdot,A)$ is
measurable for each $A\in\calS_V$. 

The first part is immediate since
$\zeta(r,\cdot)$ is a convex linear combination of two probability
measures on $\<\bar\I, \calS_V\>$, namely Dirac's  $\delta_0$
concentrated at $0$ and $m_0$. 
For the second
part, just observe that $\zeta(\cdot,A)$ is well defined for
$A\in\calS_V$ and it is a linear combination of $\calS_V$-measurable real
functions, hence $\calS_V$-measurable.
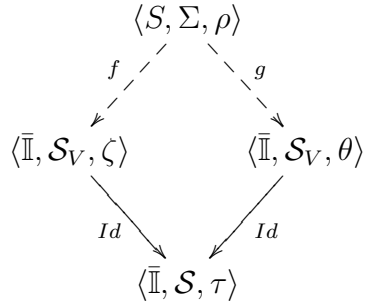
\begin{figure}[h]
\[\xymatrix@C=-1ex@R=6ex{%
 & {\<S,\Sigma,\rho\>} \ar@{-->}[dl]_f \ar@{-->}[dr]^g \\
   {\<\bar\I, \calS_V,\zeta\>} \ar[dr]_{Id} & &
   {\<\bar\I, \calS_V,\theta\>} \ar[dl]^{Id} \\
 &  {\<\bar\I, \calS,\tau\>}%
}\]
\caption{A semi-pullback.}
\label{fig:diagram}
\end{figure}

Now let $\lmp S_1:= \<\bar\I, \calS_V,\zeta\>$, $\lmp S_2:=\<\bar\I,
\calS_V,\theta\>$ and $\lmp T := \<\bar\I, \calS,\tau\>$. The identity
maps $Id : \lmp S_i \func \lmp T$ are obviously zig-zag, since $m_0$
and $m_1$ agree with $m$ over $\calS$. We will see
that there are no $\lmp S :=\<S,\Sigma,\rho\>$ and zig-zag morphisms
$f$ and $g$ that make the diagram in Figure~\ref{fig:diagram}
commutative.

If such $f$ and $g$ exist, they must be equal as functions from $S$ to
$\bar\I$ because of the commutativity of the diagram. Now let $u\in S$
such that $f(u)=0$ (recall that a zig-zag is surjective). Hence
$g(u)=0$. By the definition of zig-zag, we have: 
\[\rho(u,f^{-1}(V)) = \zeta(0,V) = m_0(V)\neq m_1(V) = \theta(0,V) = \rho(u,g^{-1}(V))\]
From this we reach a contradiction, since we have $f^{-1}(V) = g^{-1}(V)$.
\end{proof}
Given the relation between semi-pullbacks and regular conditional
probabilities (cf. \cite{Edalat}), this failure of existence of semi-pullbacks
can be traced to the fact that if $m_i(V)=0$ and $m_e(V)=1$, then
there is no regular conditional 
probability for $\tfrac{1}{2}(\overline{m} + \underline{m})$ on
$\B(\bar\I)_V$ given $\B(\bar\I)$ (see \cite[p. 81]{Breiman} and
\cite[p. 624]{Doob}).

From Theorem~\ref{t:no-semi-pullbacks} we infer that the method
of proof (i.e., the construction of a semi-pullback) used in 
\cite{DEP}  to show the logical characterization of
bisimulation cannot be applied in non-analytic spaces. It could be
argued that the existence of semi-pullbacks is not equivalent to the
transitivity of bisimulation defined as a span of zig-zags. In spite
of this, in the next section we will see that this sort of extension
of measures ensures that the transitivity of bisimilarity cannot be
proved in general.
\section{The counterexample}\label{sec:counterexample}
Following the same line of thought of the proof of Theorem~\ref{t:no-semi-pullbacks}, let $m$ be
Lebesgue measure on $\I$, let $\calS:=\B(\I)$, and let $V$ be a subset
of $\I$ that is not Lebesgue measurable. Take two extensions $m_0$ and
$m_1$ of $m$ to $\calS_V$ such that $m_0(V)\neq m_1(V)$. Let
$s,t,x\notin\I$ be mutually distinct; we may view $m_0$ and $m_1$ as measures defined on the
sum $\I\oplus\{s,t,x\}$, supported on $\I$. Recall that $\calS$ is
generated by the countable family $\calB := \{B_a:a\in \om\}$.

Let $L_3:= \om\cup\{\infty\}$. Now define a  LMP $\lmp{S_3} = \<S_3,\calS_3,
\{\tau_a  :  a\in L_3 \}\>$ such that 
\[\<S_3,\calS_3\> := \<\I\oplus\{s,t,x\},\calS_V\oplus\Pow(\{s,t,x\})\>,\] 
\begin{align*}
\tau_a(r,A) &:=\chi_{B_a}(r) \cdot \delta_x(A)\\
\tau_\infty(r,A) &:= \chi_{\{s\}}(r) \cdot m_0(A) + \chi_{\{t\}}(r) \cdot m_1(A) 
\end{align*}
when $a\in\om$ and $A\in\calS_3$.
\begin{lemma}\label{l:S3-is-lmp}
$\lmp{S_3}$ is a LMP.
\end{lemma}
\begin{proof}
We have to check that for all $l\in L_3$, $\tau_l(r,\cdot)$ is a
subprobability measure for each $r\in S_3$, and $\tau_l(\cdot,A)$ is
measurable for each $A\in\calS_3$. %

The first part follows from the fact that for all $r$, $0\leq
\chi_{B_a}(r) \leq 1$ and  $0\leq
\chi_{\{s\}}(r) + \chi_{\{t\}}(r)  \leq 1$.

For the second part, we infer measurability by the same reasoning in
the proof of Theorem~\ref{t:no-semi-pullbacks}: $\tau_l(\cdot,A)$ is
always a linear combination of measurable functions.
\end{proof}
\begin{lemma}
$s$ and $t$ are event-bisimilar.
\end{lemma}
\begin{proof}
We will check that $s$ and $t$ will not be separated by a certain  stable
\sig-algebra $\calU$. Hence they cannot be separated by the
\emph{smallest} such \sig-algebra, which is (as a relation) the \emph{greatest} event
bisimulation. 

Let $\calU:=\sig(\calB\cup\{\{s,t\},\{x\}\})$. As it easily seen from
the proof of Lemma~\ref{l:S3-is-lmp}, $\tau_a(\cdot,A)$ is
$\calU$-measurable for all $a\in \om$ and $A\in\calS_3$ (a fortiori, for
$A\in\calU$). Since $m_0$ and $m_1$ are equal on
$\sig(\B(\I)\cup\Pow(\{s,t,x\}))$, for every $A\in\calU$,
$\tau_\infty(s,A)=  \tau_\infty(t,A)$, and hence for any
$B\subseteq[0,1]$, $s$ belongs to $\tau_\infty(\cdot,A)^{-1}(B)$ if
and only if $t$ does. 
\end{proof}
\begin{theorem}\label{th:event-neq-state}
Event and state bisimilarity differ in $\lmp{S_3}$.
\end{theorem}
\begin{proof}
To prove them different it is enough to show 
that $s$ and $t$ are not state-bisimilar (and hence event bisimilarity
is not included in state bisimilarity). The strategy is simple: we
show that state bisimilarity on $\lmp{S_3}$ is the identity relation,
and hence cannot contain the pair $(s,t)$.

It is easy to show that the singleton formed with the (only) null
state $x$  must be an
$\bisim$-class. For any other $r\in S_3$, there exists $l\in L_3$ such that
$\tau_l(r,S_3)=1$ but $\tau_l(x,S_3)=0$ (and $S_3$ is obviously
$\bisim$-closed). Now take $y\neq z$ in $\I$; 
we will show that $y$ and $z$ cannot be related by $\bisim$. Since $\calB$
generates $\B(\I)$, there exists $a\in \om\subset L_3$ such that $B_a$ separates
$y$ from $z$. Without loss of generality, assume $\{y,z\}\cap B_a =
\{y\}$. Then $\tau_a(y,\{x\})= 1$ but $\tau_a(z,\{x\}) =0$. We
conclude that $\bisim$ restricted to $\I\cup\{x\}$ is the identity and (in
particular) $V\subset\I$ is $\bisim$-closed.

It remains to observe that $\tau_\infty(s,V)\neq\tau_\infty(t,V)$, and
hence $s$ and $t$ are not state-bisimilar.
\end{proof}
\begin{corollary}
The logic $\calL$ does not characterize state bisimulation for
LMP having a non-analytic base space. Moreover, the logical
characterization of state bisimulation fails for the class of LMP
having separable metrizable base spaces.
\end{corollary}
\begin{proof}
The first assertion is immediate from Theorem~\ref{th:event-neq-state}. Since
$\calS_3$ is countably generated and separates points, the 
second assertion follows from Theorem~\ref{th:event-neq-state} and
Proposition~\ref{p:sep-metriz}.
\end{proof}
It is known that in a general coalgebraic setting state bisimilarity
(defined as the existence of a span of zig-zags)
is transitive, provided the functor preserves weak pullbacks
\cite{Rutten00}. 
 By dropping alternatively states $s$ and $t$ in $\lmp{S_3}$ one may show
 that this is not the case for LMP over general measurable spaces.
\begin{corollary}\label{c:state-no-trans-span}
The relation of  bisimilarity (as given by a span of zig-zags) is
not transitive for general measurable spaces.
\end{corollary}
\begin{proof}[Sketch of proof]
Let $\lmp{S_3\setminus\{s\}} =
\<S_3\setminus\{s\},\calS_3|(S_3\setminus\{s\}),\{\tau_a  :  a\in L_3
\}\>$ be the result of ``deleting'' the state $s$ from $\lmp{S_3}$,
let $\lmp{S_3\setminus\{t\}} =
\<S_3\setminus\{t\},\calS_3|(S_3\setminus\{t\}),\{\tau_a  :  a\in L_3
\}\>$, and $\lmp{T} = 
\<S_3\setminus\{s\},\calS\oplus\Pow(\{t,x\}),\{\bar\tau_a  :  a\in L_3
\}\>$, where $\bar\tau_a$ and $\tau_a$ coincide for $a\in\om$ and for
$A\in \calS\oplus\Pow(\{t,x\})$, 
\[\bar\tau_\infty(t,A) = m(A), \qquad \bar\tau_\infty(r,A) = 0 \text{
  for $r\neq t$}\]
(note that in $\lmp{T}$ we are restricting ourselves to measurable
  subsets of the form $B\oplus X$, where $B\in\calS=\B(\I)$ and
  $X\subseteq\{t,x\}$).
The identity map $Id$ of $S_3\setminus\{s\}$ and the the map
$F:S_3\setminus\{t\} \func S_3\setminus\{s\}$ which sends $s$ to $t$
and such that $F|(\I\oplus\{x\})$ is the identity, are zig-zag morphisms $Id:
\lmp{S_3\setminus\{s\}} \func \lmp{T}$ and $F: \lmp{S_3\setminus\{t\}}
\func \lmp{T}$, respectively. Hence both $\lmp{S_3\setminus\{t\}}$ and
$\lmp{S_3\setminus\{s\}}$ are state-bisimilar to $\lmp{T}$, but they
are not state-bisimilar to each other.
\end{proof}
We can recast this last corollary in our relational framework for
bisimulation and show a serious categorical drawback of the concept of
state bisimilarity: it is not reflected by direct sum.
\begin{example}
Consider the LMP $\lmp{S_3\setminus\{s\}}$ and  $\lmp{S_3\setminus\{t\}}$
from the proof of
Corollary~\ref{c:state-no-trans-span} and let  $\lmp T'
=\<(S_3\setminus\{s,t\})\cup
\{t'\},\calS\oplus\Pow(\{t',x\}),\{\bar\tau_a  :  a\in L_3\}\>$ be the
result of renaming $t$ to $t'$ in $\lmp T$. Then $s$ and $t$ are
state-bisimilar in the sum
$\lmp U := \lmp{S_3\setminus\{t\}} \oplus \lmp{S_3\setminus\{s\}} \oplus \lmp
T'$ but in $\lmp{S_3\setminus\{t\}} \oplus \lmp{S_3\setminus\{s\}}$ they
are not.

Indeed, it is immediate that $s$, $t$ are not state-bisimilar in
$\lmp{S_3\setminus\{t\}} \oplus \lmp{S_3\setminus\{s\}}$ by using the
argument of Theorem~\ref{th:event-neq-state}. But in the sum
$\lmp{S_3\setminus\{t\}} \oplus 
\lmp{S_3\setminus\{s\}} \oplus \lmp T'$ they are. Take the equivalence
relation $R$ whose classes are $\{s,t,t'\}$ and all other triples having
corresponding elements in each of $\lmp{S_3\setminus\{t\}}$,
$\lmp{S_3\setminus\{s\}}$, and  $\lmp T'$. Then $R$ is a state
bisimulation. For this, note that if $\<U,\calU\>$ is the base space of $\lmp
U$, then every $\calU$-measurable $R$-closed subset of
$U$ must be of the form $(B\oplus B\oplus B) \cup F$, where
$B\subseteq\B(\I)$ and $F$ a finite set (in particular, $V$ cannot be the
$(\lmp{S_3\setminus\{t\}})$-part of a set in $\calU(R)$). For these sets the
transition functions behave identically.
\end{example}
Hence, state bisimilarity in LMP over general measurable spaces has
an undesirable non-local character. 
One possible conclusion of this would be to abandon state bisimilarity
and to use the event-based version, which is the main point of
\cite{coco}. But one must not overlook that the artifact of
using a non Lebesgue measurable set is rather tricky: the
Banach-Tarski Paradox, stating that a ball of radius 1 can be
decomposed in finitely many pieces that can be reassembled to form two
balls of radius 1, relies on the same device. We therefore should ask
under what circumstances we may encounter a non Lebesgue measurable
set. We discuss this in the next section.
\section{Further analysis of the construction}\label{sec:furth-analys-constr}
We know by the work of Desharnais et al.\ \cite{DEP} that in the class
of LMP over analytic
spaces, the logic $\calL$
indeed characterizes state bisimulation and hence
(obviously%
\footnote{Recall Lusin proved \cite[Theorem 21.10]{Kechris} that every
  analytic subset of $\R$ is Lebesgue measurable.}%
%
) our construction
must give a non-analytic base space. It is then natural to ask if by
imposing some regularity assumptions on the
base space we can be certain to avoid the pathological examples of
the previous sections.

Since our counterexamples need a non Lebesgue measurable subset to
start with, the first question is how complex should be the base 
measurable space $\<S,\calS\>$ as to allow non $\mu$-measurable
subsets among the sets in $\calS$. The measure of ``complexity'' we
are taking into account is the place $S$ occupies in the
\emph{projective hierarchy} of Descriptive Set Theory
\cite{Kechris,Jech_Millennium}. The first level of this hierarchy is inhabited by
analytic sets and their complements (\emph{coanalytic} or $\bPi_1^1$ sets). 
%
%
We will only be interested in the first two levels, so we give the
formal definition of the class of sets in level two and state some of
their properties.

Let $X$ be a Polish space. A subset of $X$  is in $\pca(X)$ if it is expressible
as a projection of the coanalytic set:
\begin{equation}\label{eq:1}
\pca(X)= \{\mathsf{proj}_X(C) : C \text{ coanalytic in } X\times Y,\ Y
\text{ Polish}\}.
\end{equation}
A set is in
$\bPi^1_2(X)$ if its complement is $\pca(X)$; finally define
$\Del^1_2(X):=\pca(X)\cap \bPi^1_2(X)$. We say that a measurable
space $\<S,\calS\>$ is $\pca$ (resp., $\bPi^1_2$, $\Del^1_2$) if there exists a
Polish space $X$ and $Y\in\pca(X)$ (resp.,  $\bPi^1_2(X)$,
$\Del^1_2(X)$) such that  $\<S,\calS\>\iso \<Y,\B(X)|Y\>$. All these
classes of sets  are
closed under countable unions 
and intersections and stable under restriction (if $X\subseteq Y$ are both Polish
and $\Gamma$ is  $\pca$, $\bPi^1_2$, or $\Del^1_2$, then
$\Gamma(Y)|X\subseteq\Gamma(X)$). Moreover, since the class of Polish
spaces (and their Borel spaces) are closed under  sum, this property is
inherited by  $\pca$, $\bPi^1_2$  and  $\Del^1_2$ measurable spaces.

Since every Polish space is the continuous image of the Baire space
$\Baire$, it can be proved that in Eq.~(\ref{eq:1}) we can replace $Y$
by $\Baire$. Given this preponderant role of the space of functions from $\N$ to
$\N$, recursion theory has an impact in the development of
descriptive set theory by the introduction of the \emph{lightface}
hierarchy $\Sigma_n^1$, $\Pi_n^1$ and $\Delta_n^1$; here the
notion of closed set is replaced by an effective one. We repeat the
definitions in \cite[25.1]{Jech_Millennium}.
\begin{definition}
  \begin{enumerate}
  \item A set $A\subseteq\Baire$ is $\Sigma_1^1$ if there exists a
    recursive set $R\subseteq \bigcup_{n=0}^\infty (\N^n\times
    \N^n)$ such that for all $x = (x_0,x_1,\dots)\in\Baire$,
    \[x\in A \iff \exists y \in \Baire \forall n \in \N :
    R(x|n,y|n),\]
    where $x|n := (x_0,\dots,x_{n-1})$.
  \item Let $a\in\Baire$.  A set $A\subseteq\Baire$ is $\Sigma_1^1(a)$
    ($\Sigma_1^1$ in $a$) if there exists a 
    set $R$ recursive in $a$  such that for all $x \in\Baire$,
    \[x\in A \iff \exists y \in \Baire \forall n \in \N :
    R(x|n,y|n,a|n).\]
  \item $A\subseteq\Baire$ is $\Pi_n^1$ (in $a$) if $A^\comp$ is
    $\Sigma_n^1$ (in $a$).
  \item $A\subseteq\Baire$ is $\Sigma_{n+1}^1$ (in $a$) if it is the
    projection of a  $\Pi_n^1$ (in $a$) subset of
    $\Baire\times\Baire$.
  \item $A\subseteq\Baire$ is $\Delta_n^1$ (in $a$) if $A$ is both 
    $\Sigma_n^1$ (in $a$) and  $\Pi_n^1$ (in $a$).
  \end{enumerate}
\end{definition}
We have
\[\Sig_n^1(\Baire) = \bigcup_{a\in\Baire} \Sigma_n^1(a).\]
This notions can be extended to subsets of $\R^n$ and in particular
we obtain $\Sigma_2^1(\R^n)\subseteq\Sig_2^1(\R^n)$ (as well as
$\Pi_2^1\subseteq\bPi_2^1$ and $\Delta_2^1\subseteq\Del_2^1$).

The reason for stopping at level 2 of the projective hierarchy is that
a classical result by G\"odel shows 
it is consistent with current foundations of mathematics (as given by Zermelo-Fraenkel set
theory with Choice, ZFC) that we may
find a  $\Del^1_2$ measurable space with non Lebesgue measurable sets
in its \sig-algebra. Actually, it is consistent with ZFC that there
exists a $\Delta^1_2(\R^2)$ set that is not  Lebesgue measurable.

 More precisely, G\"odel's
axiom constructibility $V=L$ (which is relative consistent with ZFC)
implies by Theorem 25.26 in Jech~\cite{Jech_Millennium} and subsequent
Corollary 25.28 that there exists  a $\Delta^1_2$ relation on $\R$
(i.e. a set $W$ in $\Delta^1_2(\R^2)$)  such that
$\<\R,W\>$ is a wellorder isomorphic to $\<\om_1,<\>$, where $\om_1$
is  the first uncountable ordinal.  And it is known that such a
relation $W$ cannot be Lebesgue measurable as a subset of $\R^2$. From
this set we will be able to  reconstruct our counterexample. 

Firstly, we manufacture a subset of $\I\times\I$ that is not
Lebesgue measurable. 
\begin{lemma}
It is consistent with ZFC that there exists a (Lebesgue) nonmeasurable subset $W'$ in  $\Del_2^1(\I\times\I)$.
\end{lemma}
\begin{proof}
By the preceding discussion, it is consistent to assume
$W\in\Del_2^1(\R^2)$ and $\<\R,W\>\iso\<\om_1,<\>$. As a consequence,
$\R$  and $\I$ are both equinumerous with $\om_1$. Define $W'$ to
be the restriction of $W$ to $\I$, i.e.,
$W'=W\cap(\I\times\I)$; we have
$W'\in\Del_2^1(\I^2)$. Again, we obtain a 
wellorder $\<\I,W'\>$ of type $\om_1$. Finally, a standard argument
shows that such a $W'$ cannot be a Lebesgue measurable subset of $\I^2$
(see, for example~\cite[Sect. 17.1]{JW}).
\end{proof}
By using this set $W'$ the construction of our counterexample can be carried
out with inessential changes. 
\begin{theorem}\label{th:unprovability}
The logical characterization of state bisimulation cannot
be proved (on ZFC basis) for the class of LMP with $\Del^1_2$ base spaces.
\end{theorem}
\begin{proof}
In the construction of Section~\ref{sec:counterexample}, we may
replace $\I$ by $\I^2$ with no trouble since $\B(\I^2)$ is likewise
countably generated (e.g. take $\calB$ to be the family of open squares with
rational vertices). Hence we now have $S:=\I^2$ and $\calS:=
\B(\I^2)$. 

Observe that $\calS\subset\Del_2^1(\I^2)$ and since 
$\Del_2^1(\I^2)$ is closed under intersection and complementation, the
sets $B\cap W'$ and $B\cap W'^\comp$ belong to $\Del_2^1(\I^2)$ for
$B\in\calS$. By Equation~(\ref{eq:sum-spaces}), 
\[\<\I^2,\calS_{W'}\> \iso \<W',\calS|W'\>\oplus  \<W'^\comp,\calS|W'^\comp\>\]
is the sum of two $\Del_2^1$ spaces, hence a
$\Del_2^1$ space it is.

The construction of $\<S_3,\calS_3\>$ will then result in a $\Del_2^1$
space since it is the sum of $\<\I^2,\calS_{W'}\>$ and a discrete space.
\end{proof}
Theorem~\ref{th:unprovability} places a limit on what can be proved in
ZFC alone. Since the Axiom of Constructibility cannot be proved in
ZFC, we only know that it is \emph{consistent} that an LMP such as
$\lmp{S_3}$ can be
constructed over a $\Del_2^1$ space. 

\section{Conclusions \& an Open Problem}\label{sec:conclusions--nice}
We constructed a LMP over a non-analytic measurable space in which state
bisimilarity and event bisimilarity differ from each other. Since the
latter is completely characterized by the modal logic $\calL$, we have a LMP such
that state bisimulation is not characterized by $\calL$. Among the
consequences of this construction we recall the non-locality of
state bisimulation: state-bisimilarity is not reflected by direct sum.

We also showed that it is consistent relative to Zermelo-Fraenkel set
theory with Choice (ZFC) that 
the logical characterization of bisimulation cannot be proved for the class of LMP
having  $\Del_2^1$ base spaces. This was accomplished by means of a
classical result of G\"odel that shows the consistency of  the
existence of a $\Del_2^1(\R^2)$ set that is not Lebesgue
measurable. $\Del_2^1$ sets lie in the second
level of the projective hierarchy (the first one being occupied by
analytic and coanalytic subsets of Polish spaces) and for uncountable
spaces, this hierarchy has $\om$ levels properly. So this cuts out the
possibility of proving the logical characterization of state
bisimulation for ``almost every'' projective space in the ZFC framework. 

We face two possibilities: to abandon state bisimulation completely in favour of the
event based one; or to consider extending our mathematical
foundations. In the second scenario, one may investigate the
consequences of the \emph{axioms of determinacy}. 
These provide a smooth theory
for the structure of analytic and projective subsets of Polish (resp.,
standard Borel) spaces 
and they are gaining wide acceptance. In particular, the axiom of 
\emph{Analytic Determinacy (AD)} (see~\cite{Kechris} for
details) implies that every set in $\pca(\R^n)$ or
$\bPi^1_2(\R^n)$ is Lebesgue measurable and enjoys various other
regularity properties.%
\footnote{From AD we also obtain regularity properties for coanalytic
  sets that are not available under ZFC: for instance, the
  \emph{perfect set property} \cite[32.2]{Kechris}.}%
\ A successful application of AD in this situation will probably depend on
presenting the problem of the logical characterization (or those
problems that imply it, like the existence of semi-pullbacks or 
regular conditional distributions) as an infinite game over an
analytic space.

A more fundamental question is
whether we actually need a non Lebesgue measurable set to furnish such
a counterexample, but we do not have an answer yet. Coanalytic sets
are Lebesgue measurable, and hence the immediate problem is to decide
whether the results in \cite{DEP,coco} can be extended to the class of
LMP with coanalytic base spaces.




\begin{ack}
I want to thank very gratefully Pedro D'Argenio and Nicol\'as
Wolovick for introducing me to the area of Markov Decision
Processes. The reason for the subindex 3 in ``$\lmp{S_3}$'' is
that this LMP completes a series of counterexamples produced while
working together on a nondeterministic version of LMP. I also
want to thank Jorge Vargas for several discussions concerning extension of
measures.  
\end{ack}
\providecommand{\noopsort}[1]{}
\begin{small}\end{small}

\bigskip

\begin{small}
\begin{quote}
CIEM --- Facultad de Matem\'atica, Astronom\'{\i}a y F\'{\i}sica 
(Fa.M.A.F.) 

Universidad Nacional de C\'ordoba --- Ciudad Universitaria

C\'ordoba 5000. Argentina.
\end{quote}
\end{small}
\end{document}